\documentclass[prl,twocolumn,showpacs,preprintnumbers]{revtex4}

\usepackage{graphicx}
\usepackage{dcolumn}
\usepackage{bm}
\usepackage{natbib}

\begin{document}


\title{Continuum Theory for Piezoelectricity in Nanotubes and Nanowires}

\author{P. J. Michalski}
\author{Na Sai}
\author{E. J. Mele}

\affiliation{Department of Physics and Astronomy, University of
Pennsylvania, Philadelphia, PA 19104}

\date{\today}

\begin{abstract}
We develop and solve a continuum theory for the piezoelectric
response of one dimensional nanotubes and nanowires, and apply the
theory to study electromechanical effects in BN nanotubes. We find
that the polarization of a nanotube depends on its aspect ratio,
and a dimensionless constant specifying the ratio of the strengths
of the elastic and electrostatic interactions. The solutions of
the model as these two parameters are varied are discussed. The
theory is applied to estimate the electric potential induced along
the length of a BN nanotube in response to a uniaxial stress.

\end{abstract}


\maketitle

Recently there has been interest in the physical properties of
materials that are obtained by wrapping a two dimensional lattice
to form a one dimensional structure. When the lateral dimensions
of the wrapped structure are reduced to the nanometer scale, the
new periodicity introduced by this wrapping can have a profound
impact on the physical properties of the system.  The III-V
boron-nitride (BN) exhibits a ground state polarization {\it and}
a piezoelectric response (electric polarization linearly coupled
to an elastic strain) when it is wrapped into a nanotube, although
the symmetry of the two dimensional BN sheet permits only a
piezoelectric response. This system thus demonstrates the
possibility of geometrical control of the piezoelectric response
of a nanomaterial, providing a new degree of freedom with
significant potential for nanoscale electromechanical
applications.

The piezoelectric response of a macroscopic three dimensional
system is well described by the continuum Landau Devonshire model
\cite{linesglass}.  However, this model does not adequately
describe the piezoelectric response of a one dimensional system
because the 1D Coulomb kernel is a short range potential (its
Fourier transform diverges at small momentum $q$ proportional to
$- \log(q)$ rather than as $1/q^2$ as in three dimensions).  Here
we show that this short range kernel leads to new physical effects
in $1$D systems such as nanotubes and nanowires.  We find that the
elastic response of a finite length piezoelectric nanotube to an
applied uniform mechanical load is in general spatially {\it
nonuniform}. The spatial variation of the electric polarization
produces a bound charge density that penetrates from the tube end
into the interior of the system. In addition, localized surface
bound charges develop exactly at the tube ends. The partitioning
of the bound charge into its ``bulk'' and ``surface'' contribution
is then determined by a competition between the electrostatic and
elastic interactions in the system. These effects are not
described in any simple model for an electrostatic depolarizing
field for the tube, as would be possible for three dimensional
samples in confined geometries (e.g. spheres, ellipsoids or thin
films), and instead require an appropriate continuum theory to
study the spatial variation of the electromechanical coupling. In
this paper we develop this model and apply it to estimate the
magnitude of experimentally accessible electromechanical effects
for the case of a BN nanotube.

We consider the effect of an applied mechanical stress on a
piezoelectric nanotube with radius $R$ and length $L$. (The
generalization to the case of a nanowire with a solid core is
straightforward.) A constant stress $\sigma(z) = f$ is applied to
the tube and it induces a strain field $\eta(z)$ where $z$ is
the coordinate along the tube axis. This strain is
linearly coupled to an electric polarization along the tube axis
through a piezoelectric constant $e_s$, so that $P(z) = e_s
\eta(z)$. In general an electric polarization of the nanotube can
be produced by the uniaxial strain (extension or compression along
the tube axis) and by tube torsion. For example, for BN nanotubes
the piezoelectric constants that couple to these two elastic
strains are determined by the chiral angle that specifies the tube
wrapping. For simplicity, in this paper we treat the case where
only one elastic degree of freedom is coupled to the polarization.
For the BN nanotubes this is realized by ``zigzag'' tubes that
have a nonzero piezoelectric coupling only to uniaxial strain, or
for ``armchair'' tubes that couple only through the tube torsion
\cite{sm03}.

The free energy of the system can be written as a sum of its bare
elastic and electrostatic contributions
\begin{equation}\label{q1}
G = G_{\rm elastic} + G_{\rm electrostatic} - \int_0^L \eta(z)
\sigma(z)\, dz.
\end{equation}
Introducing the one dimensional elastic modulus $C_1$, the elastic
energy is
\begin{equation} \label{q2}
G_{\rm elastic} = \frac{1}{2} C_1 \int_0^L \eta(z)^2 \,dz.
\end{equation}
Here $C_1 = 2 \pi R C_2$ where $C_2$ is the elastic modulus for
the two dimensional sheet (e.g. $C_{11}$ for a uniaxial strain).
$C_2$ is the relevant intensive quantity (with units energy/area)
that depends on the composition and structure of the tube, and the
one dimensional tube modulus $C_1$ is an extensive quantity
proportional to the tube radius.

Using the piezoelectric constant $e_s$ the electrostatic energy
can be expressed in terms of the spatial derivatives of the
strains as
\begin{equation} \label{q3}
G_{\rm electrostatic} = \frac{e_s^2}{2}  \int_0^L \int_0^L
\frac{\partial \eta(z)}{\partial z} V(z-z^\prime) \frac {\partial
\eta (z^\prime)}{\partial z^\prime} \,\, dz \, dz^\prime .
\end{equation}
Here $e_s$ is a {\it one dimensional} piezoelectric constant
giving the strain-induced dipole moment per unit length (with
units of charge). For BN nanotubes of moderate radius it can be
well approximated using the piezoelectric constant of an infinite
two dimensional sheet $e_2$  mapped onto the tube circumference
\cite{sm03}. Thus for most physically relevant situations $e_s
\approx 2 \pi R e_2$. $V(z-z')$ is a Coulomb kernel describing the
electrostatic interaction between rings of charge centered at
positions $z$ and $z'$ along the tube axis.

It is useful to extract the dimensional dependence of these
energies by expressing all lengths in units of the tube radius.
Then by introducing the scaled variable $\xi = z/R$, the aspect
ratio $\lambda_1 = L/R$, and another dimensionless parameter
$\lambda_2 = e_2^2/(RC_2)$, we have
\begin{eqnarray}\label{q4}
G/(\pi R^2 C_2)&=& \int_0^{\lambda_1} [(\eta (\xi))^2 - 2\beta \eta (\xi)] \,
d\xi \nonumber
\\& + & {2\pi \lambda_2} \int_0^{\lambda_1} \!\!\! \int_0^{\lambda_1} \frac{\partial
\eta(\xi)}{\partial \xi} V(\xi-\xi^\prime)\frac{\partial
\eta(\xi^\prime)}{\partial \xi^\prime}\,d\xi\,d\xi^\prime  .
\end{eqnarray}
Here $\beta = f/(2 \pi R C_2)$ is the equilibrium strain in the
absence of any electrostatic coupling. The competition between the
electrostatic and elastic effects on the tube is therefore
determined by the relative sizes of the scaled lengths $\lambda_1$
and $\lambda_2$.  In the regime where $\lambda_2 \ll \lambda_1$
the electrostatic interactions are negligible, and the equilibrium
strain is nearly constant along the length of the tube, albeit
with a reduction near the tube ends. Conversely when $\lambda_2
\gg \lambda_1$, the electrostatic interactions dominate and the
system seeks to minimize the spatial derivatives of the strain,
subject to the constraints of mechanical equilibrium. This leads
to a situation where the bound charge density penetrates deep into
the tube interior and the applied mechanical stress is nearly
completely balanced by the electrostatic interactions. Equation
(\ref{q4}) also demonstrates that because of the scaling of
the one-dimensional modulus with tube radius, the induced strains
are inversely proportional to $R$, i.e. for a fixed aspect ratio
large radius tubes are mechanically ``stiffer''.

The equilibrium strain field is obtained by minimizing $G$ and
satisfies
\begin{equation}\label{q5}
\eta (\xi) = \beta + 2 \pi \lambda_2 \int_0^{\lambda_1}
\frac{\partial V(\xi-\xi^\prime)}{\partial \xi} \frac{\partial
\eta (\xi^\prime)}{\partial \xi^\prime} \, d\xi^\prime .
\end{equation}
Possible stepwise discontinuities of $\eta(z)$ at the sample
boundaries can be treated by explicitly writing
\begin{equation}\label{q6}
\frac{\partial \eta (\xi^\prime)}{\partial \xi^\prime} = \eta (0)
\delta (\xi^\prime) - \eta (\lambda_1) \delta (\xi^\prime
-\lambda_1) + \frac{\partial \eta_i (\xi^\prime)}{\partial
\xi^\prime},
\end{equation}
where $\eta_i$ denotes the strain field in the sample interior.
Then, by integrating Eq. (\ref{q5}) by parts we get
\begin{equation}\label{q7}
\eta_i (\xi) = \beta + 2 \pi \lambda_2 \int_0^{\lambda_1}
\frac{\partial^2 V(\xi-\xi^\prime)}{\partial \xi^{\prime 2}}
\eta_i (\xi^\prime) \, d\xi^\prime  .
\end{equation}
In the following we will drop the subscript $i$, and the
$\xi^\prime$ integrals are understood to extend over the interior
of the tube.

The solution of Eq. (\ref{q7}) requires a specification of an
appropriate Coulomb kernel $V(z)$. In the simplest continuum model
where $V(z)$ describes the interaction of rings of charge, we have
\begin{equation}\label{q8}
V(\xi-\xi^\prime)=\frac{1}{2\pi} \int_0^{2\pi}
\frac{d\theta}{\sqrt{(\xi-\xi^\prime)^2 + \sin^2(\theta) +
(1-\cos(\theta))^2}}.
\end{equation}
This kernel diverges logarithmically at short range. Thus in this
model a surface localized bound charge density $\rho(0) \delta(z)$
at the tube end leads to a divergent electrostatic energy.
Therefore the boundary condition for the strain field in this
model becomes $\eta(0) = \eta(L) = 0$. This infinite energy is an
artifact of the model and signals a breakdown of the continuum
theory at short length scales. To avoid this difficulty, we
instead use a ``softened'' kernel
\begin{equation}\label{q9}
V_{\rm soft} (\xi-\xi^\prime) = \frac{1}{|\xi-\xi^\prime| +
\alpha},
\end{equation}
which retains the correct long range behavior but remains finite
for $\xi \rightarrow \xi'$. This softened kernel allows for the
presence of a nonzero surface charge density $e_s \eta(0)
\delta(z)$ at the ends the tube.  The value of $\alpha$ determines
the energy ``cost'' of surface charges, and therefore controls the
value of $\eta (0)$. Ultimately the distribution of the bound
charge into its ``surface'' and ``bulk'' contributions is
determined by a self consistent solution to the equation of state
(\ref{q7}).

Once the equilibrium charge density is determined the electric
potential at position $z$ along the tube is obtained by evaluating
\begin{equation} \label{q10}
U(z) = \int_0^L V(z-z^\prime) \rho(z^\prime) \, dz^\prime .
\end{equation}
After integrating by parts and transforming to scaled variables
the potential is expressed
\begin{equation}\label{q11}
U(\xi) = 2 \pi e_2 \int_0^{\lambda_1} \frac{\partial
V(\xi-\xi^\prime)}{\partial \xi^\prime} \eta(\xi^\prime) \, d
\xi^\prime  .
\end{equation}
Since the equilibrium condition in Eq. (\ref{q7}) requires that
\begin{equation}\label{q12}
\eta(\xi) - \beta = - \frac{\lambda_2}{e_2} \frac{\partial
U(\xi)}{\partial \xi}
\end{equation}
the potential can be expressed as an integral over the strain
field $\eta$, and we obtain
\begin{equation}\label{q13}
U(\xi) = \frac{e_2}{\lambda_2} \int_{\xi}^{\lambda_1/2}
(\eta(\xi^\prime) - \beta) \, d \xi^\prime ,
\end{equation}
where we have set the electric potential to zero at the center of
the tube.

\begin{figure}
\includegraphics[width = 86mm]{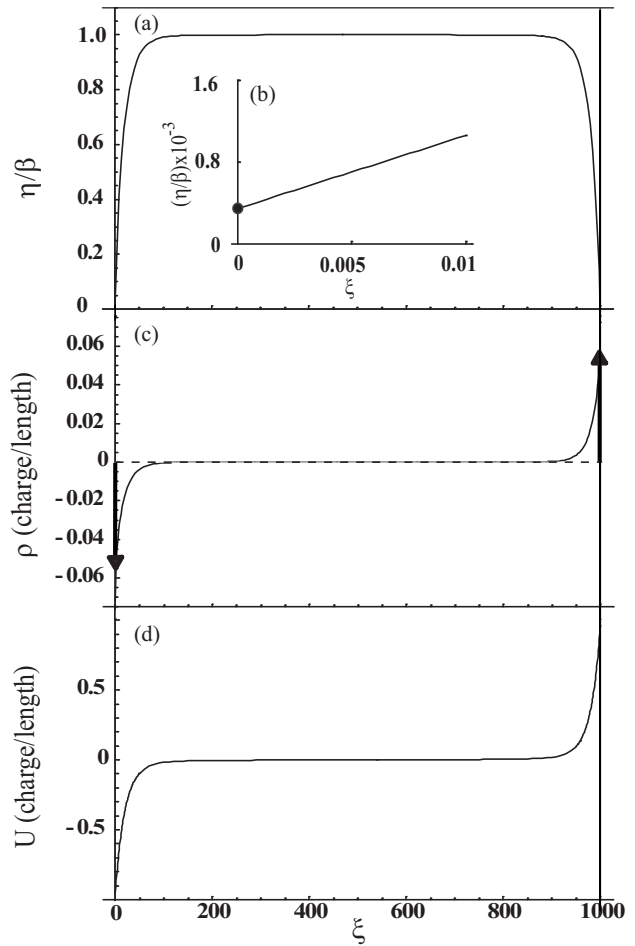}
\caption{\label{fig:eregime} Equilibrium solutions for a
piezoelectric tube in the elastically dominated limit.  (a)
Equilibrium strain profile from Eq. (\ref{q7}). The strain is well
approximated by the elastic limit over most of the tube.  (b)
Inset, strain value at tube boundary. Notice $\eta (0) \not= 0$
with our choice of $V$. (c) The bound charge density, $\rho = -e_s
\partial \eta/
\partial z$.  The
arrows represent delta functions in the charge density. (d)
Equilibrium potential from Eq. (\ref{q13}). The graphs were
generated using $\beta = 1$, $\lambda_1 = 1000$, $\lambda_2 =
10/\pi$, and $e_{s} = 1 $ (charge). }
\end{figure}

\begin{figure}
\includegraphics[width = 86mm]{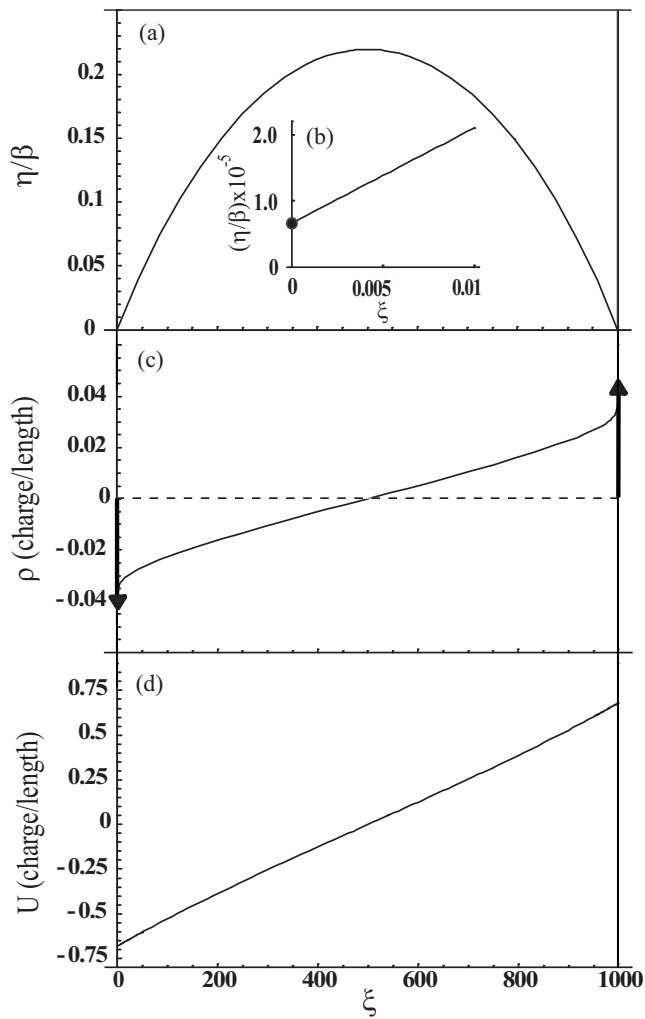}
\caption{\label{fig:cregime} The equilibrium solutions for a
piezoelectric tube in the electrostatically dominated limit. (a)
Equilibrium strain profile from Eq. (\ref{q7}). Notice that the
maximum value of the strain is about 1/5 of the elastic limit.
(b) Inset, strain value at tube boundary.  Notice $\eta (0) \not=
0$ with our choice of $V$. (c) The bound charge density, $\rho =
-e_s
\partial \eta/
\partial z$.  The arrows represent delta functions in the charge density. (d) Equilibrium potential from Eq.
(\ref{q13}). The graphs were generated using $\beta = 1$,
$\lambda_1 = 1000$, $\lambda_2 = 10000/\pi$, and $e_{s} = 32$
(charge).}
\end{figure}

The equilibrium solutions for the strain field $\eta(z)$ (Eq.
\ref{q7}), the bound charge density $\rho=-e_s
\partial \eta/\partial z$ and the electrostatic potential $U(z)$
(Eq. \ref{q13}) are displayed in Fig. (\ref{fig:eregime}) for the
model in the regime where the elastic interactions dominate
($\lambda_2 \ll \lambda_1$) and in Fig. (\ref{fig:cregime}) for
the regime where the electrostatic interactions dominate
($\lambda_1 \ll \lambda_2$).  The softened kernel was used with
$\alpha =.001$.  The data in these plots was obtained by finite
element analysis. The tube length was divided into $1000$
unequally spaced bins ($1/3$ evenly spaced in the first $10\%$ of
the tube, $1/3$ evenly spaced in the last $10\%$, and $1/3$ evenly
spaced across the middle $80 \%$ of the tube). Bin number was
increased in steps of $100$ until successive solutions differed by
less than $1.2\%$. The slowest convergence occurs at the end of
the tubes, with the solution differing by less than $.2\%$ on the
last iteration over the middle $99.8\%$ of the tube.

In the elastically dominated limit, the equilibrium strain is
nearly constant in the interior region of the tube where $\eta
\approx \beta$.  The electrostatic interactions effectively
increase the elastic moduli near the surface region, leading to a
suppression of the strain.  This suppression implies a spatial
variation of the polarization and a volume bound charge density
that is obtained self consistently by solving Eq. (\ref{q7}).
Finally, we see that the equilibrium strain is in general {\it
nonzero} at the tube ends. Thus, in addition to the continuous
volume bound charge density, a true surface localized bound charge
is obtained at the terminations. In a long (semi-infinite) tube
the equilibrium strain asymptotically relaxes to its elastic limit
following a power law, e.g. $1 - \eta(\xi)/\beta \propto
(\gamma/\xi)^2$ where $\gamma = \sqrt{2 \pi \lambda_2}$. The
distribution of the bound charge is plotted in the central panel,
which shows that the main effect of the electrostatic interaction
is to ``spread'' the bound charge density near the tube ends. In
the elastically dominated limit most of the potential drop across
the tube, shown in Fig. (\ref{fig:eregime}-d), occurs in the
regions near the tube ends.

This behavior is contrasted with the situation for the
electrostatically dominated limit shown in Fig.
(\ref{fig:cregime}). In this regime, the system seeks to reduce
its strain gradients subject to the constraint of mechanical
equilibrium.  This leads to a state in which the bound charge
extends along the entire length of the tube. The numerical
solution for the equilibrium strain shown in Fig.
(\ref{fig:cregime}-a) is reasonably well approximated using only
the lowest allowed Fourier component with the ends of the tube
clamped. The overall magnitude of the strain is also strongly
reduced, since it is limited by the electrostatic coupling
strength instead of the bare elastic constant.  The bound charge
density vanishes at the center of the tube by symmetry and varies
approximately linearly as a function of position, with logarithmic
corrections close to the ends of the tube. Interestingly, Eq.
(\ref{q12}) requires that the electric potential is a linear
function of position in the limit that $\eta(z) \ll \beta$, as is
shown in Fig. (\ref{fig:cregime}-d).

Since the one dimensional Coulomb kernel is a short range
interaction it is useful to consider the limit where the kernel is
a contact interaction $V(\xi-\xi^\prime)=\delta (\xi-\xi^\prime)$.
Then the free energy in Eq. (\ref{q4}) reads
\begin{equation} \label{q14}
G/(\pi R^2 C_2) = \int_0^{\lambda_1}\!\!\left[(\eta (\xi))^2 - 2\beta \eta
(\xi)+ 2\pi \lambda_2 \left(\frac{\partial \eta(\xi)}{\partial
\xi}\right)^2 \right]\,d\xi .
\end{equation}
Up to an additive constant, this describes the free energy of a
superconducting disk, with penetration depth
$\sqrt{2\pi\lambda_2}$ immersed in a perpendicular field $H_0 =
-\beta$.  The equilibrium solution to this model is obtained
analytically as
\begin{equation}\label{q15}
\eta(\xi)=\beta\left(1-\frac{\cosh ((\xi-\lambda_1 /
2)/\sqrt{2\pi\lambda_2}\,)}{\cosh ((\lambda_1 /2) /
\sqrt{2\pi\lambda_2}\,)}\right)  .
\end{equation}
Thus the strain profile in the elastically dominated limit shown
in Fig. (\ref{fig:eregime}-a) give the piezoelectric analog to the
Meissner expulsion of an applied field (stress) and the
electrostatic regime of Fig. (\ref{fig:cregime}-a) corresponds to
the limit of a small superfluid density where the penetration
depth is of order the length of the system.  For the contact
interaction model the induced potential is simply proportional to
the bound charge density, so that a measurement of the spatial
variation of the potential probes the piezoelectric analog of the
London penetration depth.  Note however that the physical kernel
retains an asymptotic $1/\xi$ tail leading to a power law decay of
the strain field.

Recently, BN nanotubes were predicted to exhibit a significant
piezoelectic response \cite{mk02, ncmbn03}.  Here we predict the
piezoelectric response of a ``zigzag'' BN nanotube under uniform
longitudinal tensile or compressive stress. The 2D elastic
constants for various chiral BN tubes were calculated in
Ref.~\onlinecite{hgbr98}; here we use an average value of
$C_{2}=.30$ TPa$\,\,$nm . Ref.~\onlinecite{sm03} gives the
piezoelectric constant for a BN sheet as $e_{2}=.12$ e/Bohr.
Experimentally accessible forces are on the nano-Newton scale; we
arbitrarily assume a force of $1$ nN. Typical tube dimensions are
$R=1$ nm and $L=1$ $\mu$m.

Using these values we find that $\lambda_1 = 1000$, $\lambda_2 =
4.4 \times 10^{-3}$, and $\beta = 5.3 \times 10^{-4}$. These
values give an asymptotic decay constant of $\gamma = .17$,  and the
strain to relaxes to its elastic limit very quickly. Our numerical
solution shows the equilibrium strain is within $1\%$ of the
elastic limit at a depth of $3R$. Thus the bound charge density,
although continuous, barely penetrates the bulk of the tube, and
the entire potential drop occurs near the ends. Applying the
continuum theory to this system and converting to SI units we
obtain an estimate of $.48$ V across the tube for an applied
tension $f=1 \, {\rm nN}$. It would be interesting to
experimentally measure this strain induced electrostatic potential
and to resolve its spatial variation near the tube ends using
scanning probe potentiometry.

The theory for a solid core nanowire follows the same formalism,
but unlike the nanotube, in a nanowire the corresponding
$\lambda_2$ is fixed entirely by the material parameters, {\it
independent} of $R$.  This occurs for a nanowire because the
extensive variables, $e_s$ and $C_1$, are proportional to the
cross-sectional {\it area} rather than the circumference.  They
are related to the relevant intensive variables by $e_s = \pi R^2
e_3$ (where $e_3$ is the 3D piezoelectric constant in units of
charge/area) and $C_1 = \pi R^2 C_3$ (where $C_3$ is the 3D
elastic constant in units of pressure).  The equilibrium strain
then satisfies Eq. (\ref{q7}) with $\lambda_2 = e_3^2/(2 C_3)$ and
$\beta = f/(\pi R^2 C_3)$.  Thus two nanowires with the same
aspect ratio but different radii will have the same scaled strain,
$\eta /\beta$, unlike hollow nanotubes where larger radius tubes
have smaller $\lambda_2$ and thus are elastically dominated.

This work was supported by the Department of Energy under grant
DE-FG$02$-ER$0145118$ and by the National Science Foundation under
grant DMR-$00$-$79909$.

\end{document}